\documentclass[aps,twocolumn,superscriptaddress]{revtex4}
\usepackage{graphicx}

\newcommand{\bra}[1] {\left\langle #1 \right|}
\newcommand{\ket}[1] {\left| #1 \right\rangle}
\newcommand{\avg}[1] {\left\langle #1 \right\rangle}

\begin{document}

\title{How the result of a measurement of a component of the spin
of a spin-1/2 particle can turn out to be 100 without using weak
measurements}

\author{S. Ashhab}
\affiliation{Advanced Science Institute, The Institute of Physical
and Chemical Research (RIKEN), Wako-shi, Saitama 351-0198, Japan}
\affiliation{Physics Department, Michigan Center for Theoretical
Physics, The University of Michigan, Ann Arbor, Michigan
48109-1040, USA}

\author{Franco Nori}
\affiliation{Advanced Science Institute, The Institute of Physical
and Chemical Research (RIKEN), Wako-shi, Saitama 351-0198, Japan}
\affiliation{Physics Department, Michigan Center for Theoretical
Physics, The University of Michigan, Ann Arbor, Michigan
48109-1040, USA}

\date{\today}



\begin{abstract}
We discuss two questions related to the concept of weak values as
seen from the standard quantum-mechanics point of view. In the
first part of the paper, we describe a scenario where unphysical
results similar to those encountered in the study of weak values
are obtained using a simple experimental setup that does not
involve weak measurements. In the second part of the paper, we
discuss the correct physical description, according to quantum
mechanics, of what is being measured in a weak-value-type
experiment.
\end{abstract}

\maketitle

\section{Introduction}

The first part of the title of this paper (all but the last four
words) is taken from the title of a paper written by Aharonov,
Albert and Vaidman (AAV) over twenty years ago \cite{Aharonov}. In
that paper AAV introduced the concept of weak values. This concept
immediately caused controversy \cite{Comments}, but over the years
it has proved to be a useful paradigm for considering questions
related to quantum measurement and the foundations of quantum
mechanics. For example, the observation of paradoxical values in a
weak-value-type measurement has been linked to the violation of
the Leggett-Garg inequality, which can be used to test realism
\cite{LeggettGarg,Williams,Romito}.

In the setup considered by AAV, a beam of spin-1/2 particles
propagates through a non-uniform magnetic field in a
Stern-Gerlach-type experiment, where the trajectory of a given
particle is affected by the spin state of the particle. The
modification from the original Stern-Gerlach experiment is that,
in the path of its propagation, the beam encounters two regions in
space with magnetic fields. The magnetic field gradient in the
first region is designed such that it creates a tendency for
particles whose $x$-component of the spin (which we denote by
$S_x$) is positive to develop a finite component of the momentum
in the positive $x$ direction and for particles whose $S_x$ is
negative to develop a finite component of the momentum in the
negative $x$ direction. After exiting this region in space, the
beam enters a second region where a $z$-component in the momentum
develops based on the $z$-component of the spin ($S_z$). Either
one of these stages would constitute a measurement of the spin
along some direction: by setting up a screen that the beam hits
sufficiently far from the field-gradient region, the position
where a given particle hits the screen serves as an indicator of
the particle's spin state. When combined, they create a situation
where two non-commuting variables are being measured in
succession. If (1) the first measurement stage is designed to be a
weak measurement, (2) the particles in the beam are created in a
certain initial state [e.g.~close to being completely polarized
along the positive $z$-axis] and (3) only those particles for
which the second measurement produces a certain outcome [in this
example, a negative $z$-component of the spin], then the average
value of the spin's $x$-component indicator can suggest values of
this component of spin being much larger than 1/2, a situation
that seems paradoxical.

A number of studies have already pointed out that since in the AAV
setup two non-commuting variables are being measured in
succession, quantum mechanics forbids treating them as independent
measurements whose outcomes do not affect one another
\cite{Comments}. In this paper we start by presenting an example
that demonstrates the role of interpretation in obtaining
unphysical results in a weak-measurement-related setup. The setup
is chosen to be very simple in order to remove any complications
in the analysis related to the successive measurement of
non-commuting variables. In the second part of the paper, we
present the proper analysis (from the point of view of quantum
mechanics) of the measurement results obtained in an AAV setup.

\section{Question 1: Unphysical results of the AAV type in an alternative setup}

Let us consider the following situation: An experimenter purchases
a device for measuring the $z$-component of a spin-1/2 particle.
The device produces one of two readings, 0 or 1. The experimenter
goes to the lab and calibrates the device. The calibration is done
by preparing $10^6$ particles in the spin up state, measuring them
one by one, and then doing the same for the spin down state. Let
us say that the result of the calibration procedure is that for
the spin up state the device shows the reading ``1'' in 50.25\% of
the experimental runs and the reading ``0'' in 49.75\% of the
runs. For the spin down state, the probabilities are reversed.
Clearly, the reading of the measurement device is only weakly
correlated with the spin state of the measured particle. The
experimenter takes this fact into account and reaches the
following conclusion: If I have a large number of identically
prepared spin-1/2 particles and measure them using this device, I
will obtain a probability for the reading ``1''. Using the results
of the calibration procedure, the expectation value of the spin
$z$-component for the prepared state will be given by the formula:
\begin{equation}
\avg{S_z} = \left({\rm Prob_1}-0.5\right) \times 200.
\end{equation}
If the probability of obtaining the outcome ``1'' is 0.5025, the
above formula gives 1/2. If the probability of obtaining the
outcome ``1'' is 0.4975, the above formula gives -1/2. It looks
like the device is ready to be used. The experimenter now performs
an experiment that involves, as its final step, a measurement of
$S_z$. Surprisingly, the measurement device shows the reading
``1'' every time the experiment is repeated, leading the
experimenter to conclude that the value of the spin is in fact
100. Thus one has a paradox.

The resolution of the paradox in the above story lies in the fact
that the device was not a weak-measurement device as the
experimenter assumed, but a strong-measurement device whose
reading is perfectly correlated with the spin state of the
measured particle. The only problem is that at some point before
the measurement device was calibrated, its spin-sensing part was
rotated from being parallel to the $z$-axis to an axis that makes
an angle 89.7135 with the $z$-axis (note here that $\cos^2
(89.7135/2) \approx 0.5025$). Not surprisingly, the calibration
procedure produced the probabilities 0.5025 and 0.4975. In the
``real'' experiment, the spins were all aligned with the
measurement axis of the device, and the reading ``1'' was observed
in all the runs. The paradox is therefore resolved.

An unquestioning believer in quantum mechanics might say that the
situation discussed in Ref.~\cite{Aharonov} has a large amount of
overlap with the story presented above. In both cases a perfectly
acceptable measurement is performed. The reason for obtaining a
paradoxical measurement result is simply the wrong interpretation
of what the measurement device is measuring and the resulting
erroneous mapping from measurement outcomes to values of the
measured quantity.

\section{Question 2: Correct explanation of results in an AAV setup}

\begin{figure}[h]
\includegraphics[width=6.0cm]{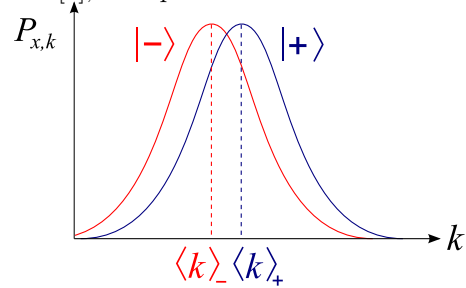}
\caption{(color online) Schematic diagram of the probability
distributions of the possible measurement outcomes of a weak
measurement (labelled by the index $k$) for the two states of the
measurement basis, $\ket{+}$ and $\ket{-}$.}
\end{figure}

We now turn to the question of the correct interpretation of the
AAV experiment according to quantum mechanics. Instead of the
original, Stern-Gerlach-type experiment analyzed by AAV, we
formulate the problem slightly differently. We consider a spin-1/2
particle that is subjected to two separate measurements. As a
first step, a weak measurement is performed in the basis $\left\{
\ket{+},\ket{-} \right\}$, where $\ket{\pm} = (\ket{\uparrow} \pm
\ket{\downarrow})/\sqrt{2}$ and the states $\ket{\uparrow}$ and
$\ket{\downarrow}$ are the eigenstates of $\hat{S}_z$. This
measurement can produce any one of a large number of possible
outcomes, with probability distributions as shown in Fig.~1. This
measurement constitutes a weak measurement of $\hat{S}_x$. As
discussed in \cite{Ashhab}, each possible outcome is associated
with a measurement matrix $\hat{U}_{x,k}$, where the index $k$
represents the outcome that is observed in a given run of the
experiment. If the outcome $k$ occurs with probability $P_{x,k}$
for the system's maximally mixed state, i.e.~when averaged over
all possible initial states, and it provides measurement fidelity
$F_{x,k}$ (in favor of the state $\ket{+}$), the measurement
matrix $\hat{U}_{x,k}$ is given by
\begin{widetext}
\begin{eqnarray}
\hat{U}_{x,k} & = & \sqrt{P_{x,k}} \left\{ \sqrt{1+F_{x,k}}
\ket{+}\bra{+} + \sqrt{1-F_{x,k}} \ket{-}\bra{-} \right\} \nonumber \\
& = & \frac{\sqrt{P_{x,k}}}{2} \left( \begin{array}{cc}
\sqrt{1+F_{x,k}} + \sqrt{1-F_{x,k}} & \sqrt{1+F_{x,k}} - \sqrt{1-F_{x,k}} \\
\sqrt{1+F_{x,k}} - \sqrt{1-F_{x,k}} & \sqrt{1+F_{x,k}} + \sqrt{1-F_{x,k}} \\
\end{array}
\right).
\end{eqnarray}
\end{widetext}
We shall use the convention where a measurement that favors the
state $\ket{-}$ has a negative value of $F_{x,k}$ and
$\hat{U}_{x,k}$ is given by the same expression as above. It is
worth mentioning here that the overall, or average, fidelity of
this measurement can be obtained by averaging over all possible
initial states and all possible outcomes:
\begin{equation}
\avg{F_x} = \sum_k P_{x,k} \left| F_{x,k} \right|.
\end{equation}
After the weak $x$-basis measurement, a strong measurement in the
$\left\{ \ket{\uparrow},\ket{\downarrow} \right\}$ basis is
performed. This strong measurement step can be described by two
outcomes with corresponding measurement matrices
\begin{eqnarray}
\hat{U}_{z,1} & = & \ket{\uparrow}\bra{\uparrow} \nonumber \\
& = & \left( \begin{array}{cc}
1 & 0 \\
0 & 0 \\
\end{array}
\right) \nonumber \\
\hat{U}_{z,2} & = & \ket{\downarrow}\bra{\downarrow} \nonumber \\
& = & \left( \begin{array}{cc}
0 & 0 \\
0 & 1 \\
\end{array}
\right).
\end{eqnarray}

As mentioned above, paradoxes arise if one treats the $x$-basis
and $z$-basis measurements as two separate measurements that
provide complementary information. Instead, one should treat each
pair of outcomes as a single \textit{combined-measurement}
outcome. The maximum amount of information in a given run of the
experiment can be extracted as follows \cite{Ashhab}: given that
the outcome pair $\{k,l\}$ was observed, one can construct the
combined-measurement matrix
\begin{equation}
\hat{U}_{{\rm Total},k,l} = \hat{U}_{z,l} \hat{U}_{x,k}.
\end{equation}
From the matrices $\hat{U}_{{\rm Total},k,l}$ one can construct a
so-called positive operator-valued measure (POVM) defined by the
matrices $\hat{M}_{k,l}$:
\begin{equation}
\hat{M}_{k,l} = \hat{U}_{{\rm Total},k,l}^{\dagger} \hat{U}_{{\rm
Total},k,l},
\end{equation}
where the superscript $\dagger$ represents the transpose conjugate
of a matrix. In particular,
\begin{widetext}
\begin{eqnarray}
\hat{M}_{k,1} & = & \frac{P_{x,k}}{2} \left( \begin{array}{cc}
\left( \sqrt{1+F_{x,k}} + \sqrt{1-F_{x,k}} \right)^2 & 2 F_{x,k} \\
2 F_{x,k} & \left( \sqrt{1+F_{x,k}} - \sqrt{1-F_{x,k}} \right)^2 \\
\end{array}
\right) \nonumber \\
& = & P_{x,k} \left( 1 + \ket{\psi_{k,1}}\bra{\psi_{k,1}} -
\ket{\overline{\psi}_{k,1}}\bra{\overline{\psi}_{k,1}} \right) \nonumber \\
& = & 2 P_{x,k} \ket{\psi_{k,1}}\bra{\psi_{k,1}}
\end{eqnarray}
\end{widetext}
where
\begin{eqnarray}
\ket{\psi_{k,1}} & = & \cos\frac{\theta_{k}}{2} \ket{\uparrow} +
\sin\frac{\theta_{k}}{2} \ket{\downarrow} \nonumber \\
\ket{\overline{\psi}_{k,1}} & = & \sin\frac{\theta_{k}}{2}
\ket{\uparrow} - \cos\frac{\theta_{k}}{2} \ket{\downarrow} \nonumber \\
\sin\theta_{k} & = & F_{x,k}.
\end{eqnarray}
Similarly one can find that
\begin{eqnarray}
\hat{M}_{k,2} & = & 2 P_{x,k} \ket{\psi_{k,2}}\bra{\psi_{k,2}}
\nonumber \\
\ket{\psi_{k,2}} & = & \sin\frac{\theta_{k}}{2} \ket{\uparrow} +
\cos\frac{\theta_{k}}{2} \ket{\downarrow},
\end{eqnarray}
with $\theta_k$ given by the same expression as above.

As discussed in Ref.~\cite{Ashhab}, one can obtain the measurement
basis and fidelity that correspond to the outcome defined by
$\{k,l\}$ by diagonalizing the matrix $\hat{M}_{k,l}$. Since
$\hat{M}_{k,l}$ is a hermitian matrix, its two eigenvalues
($m_{k,l,1}$ and $m_{k,l,2}$, with $m_{k,l,1} \geq m_{k,l,2}$)
will be real and its two eigenstates ($\ket{\psi_{k,l}}$ and
$\ket{\overline{\psi}_{k,l}}$) will be orthogonal quantum states
that define a basis (the measurement basis). Note that because the
second measurement in the problem considered here is a strong
measurement, we always have $m_{k,l,2}=0$.

The different outcomes produce different measurement bases, thus
this measurement cannot be thought of in the usual sense of
measuring $S_{\bf n}$ with ${\bf n}$ being some fixed direction.
Therefore, the measurement basis is determined stochastically for
each (combined) measurement (note that after the strong $z$-basis
measurement, the system always ends up in one of the states
$\{\ket{\uparrow},\ket{\downarrow}\}$, even though the
combined-measurement basis can be different from the basis
$\{\ket{\uparrow},\ket{\downarrow}\}$). By analyzing all the
measurement data, one can perform partial quantum state tomography
and determine the $x$ and $z$-components in the initial state of
the system (assuming of course that all copies are prepared in the
same state, which can be pure or mixed). Note that in this setup
no information about $S_y$ can be obtained from the measurement
outcome.

We now ask whether information can be extracted from the $x$-basis
and $z$-basis measurements separately, i.e.~by disregarding the
outcome of one of the two measurement steps. The answer is yes,
provided care is taken in interpreting the results. Extracting an
$x$-basis measurement from a given measurement outcome is
straightforward. All one has to do is disregard the outcome of the
$z$-basis measurement, since this measurement is performed after
the $x$-basis measurement and cannot affect the outcome of the
$x$-basis measurement. Therefore, by disregarding the outcome of
the $z$-basis measurement, one obtains an $x$-basis measurement
with overall fidelity $\avg{F_x}$. The situation is somewhat
trickier if one wants to extract a $z$-basis measurement from the
measurement outcome. One can disregard the outcome of the
$x$-basis measurement, but one must take into account the fact
that this measurement generally changes the state of the system
before the $z$-basis measurement is performed. The effect of the
$x$-basis measurement is to reduce the fidelity of the $z$-basis
measurement. One can calculate this reduced fidelity as follows:
Let us assume that the system starts in the initial state
$\ket{\uparrow}$. After the $x$-basis measurement is performed and
the outcome $k$ (with fidelity $F_{x,k}$) is observed, the state
of the system is transformed into a new pure state $\ket{\psi_{\rm
int}}$ with $\left| \bra{\psi_{\rm int}} \hat{\sigma}_x
\ket{\psi_{\rm int}} \right| = F_{x,k}$. Since
\begin{equation}
\left| \bra{\psi_{\rm int}} \hat{\sigma}_x \ket{\psi_{\rm int}}
\right|^2 + \left| \bra{\psi_{\rm int}} \hat{\sigma}_y
\ket{\psi_{\rm int}} \right|^2 + \left| \bra{\psi_{\rm int}}
\hat{\sigma}_z \ket{\psi_{\rm int}} \right|^2 = \frac{1}{4}
\end{equation}
for any pure state and here we have $\left| \bra{\psi_{\rm int}}
\hat{\sigma}_y \ket{\psi_{\rm int}} \right| = 0$, we find that
after the $x$-basis measurement $4 \left| \bra{\psi_{\rm int}}
\hat{\sigma}_z \ket{\psi_{\rm int}} \right|$ is reduced from 1 to
$\sqrt{1-F_{x,k}^2}$. If $F_{x,k}$ is independent of $k$, one
obtains the relation (in this context, see
e.g.~Ref.~\cite{Kurotani})
\begin{equation}
\avg{F_x}^2 + \avg{F_z}^2 = 1.
\end{equation}

We now take one final look at the AAV gedankenexperiment. We
choose a specific form for the $x$-basis measurement, which is
essentially the same one used by AAV
\begin{eqnarray}
P_{x,k} & = & \frac{1}{\sqrt{2\pi k_{\rm rms}^2}} \exp \left\{
-\frac{k^2}{2k_{\rm rms}^2} \right\}
\nonumber \\
F_{x,k} & = & \sqrt{\frac{\pi}{2}} \frac{\avg{F_x}}{k_{\rm rms}}
k,
\end{eqnarray}
with $k$ running over all integers from $-\infty$ to $+\infty$ and
$k_{\rm rms}$ assumed to be a large number. Note that the above
expression violates the constraint that $F_{x,k}<1$. However,
provided that $\avg{F_x} \ll 1$, the above expression can be
treated as a good approximation of the realistic situation for all
practical purposes. A simple calculation shows that in this case
\begin{eqnarray}
\avg{F_z} & = & \sum_{k=-\infty}^{\infty} \sqrt{1-F_{x,k}^2}
P_{x,k}
\nonumber \\
& \approx & 1 - \frac{\pi \avg{F_x}^2}{4},
\end{eqnarray}
such that
\begin{equation}
\avg{F_x}^2 + \avg{F_z}^2 \approx 1 - \frac{\pi - 2}{2}
\avg{F_x}^2.
\end{equation}

If the measured system is prepared in one of the states
$\ket{\pm}$, the average value of $k$ that is obtained in an
ensemble of measurements (all with the same initial state) is
\begin{equation}
\avg{k}_{\ket{\pm}} = \pm \frac{\avg{F_x} k_{\rm rms}}{2}.
\end{equation}
The small difference between $\avg{k}_{\ket{+}}$ and
$\avg{k}_{\ket{-}}$ is the reason why the $x$-basis measurement
qualifies as a weak measurement of $S_x$. We now consider the full
measurement procedure. If one prepares the measured system in a
state that is very close to $\ket{\uparrow}$, most $z$ basis
measurements will produce the outcome $l=1$. Only a small fraction
of the experimental runs will produce the outcome $l=2$. If the
initial state deviates slightly from $\ket{\uparrow}$, i.e.
\begin{equation}
\ket{\psi_i} = \cos\frac{\alpha}{2} \ket{\uparrow} +
\sin\frac{\alpha}{2} \ket{\downarrow},
\end{equation}
then outcomes with negative values of $k$ and $l=2$ will be
suppressed the most (assuming $\alpha$ is positive), because these
outcomes correspond to states that are orthogonal or almost
orthogonal to the initial state (making their occurrence
probabilities particularly small). One therefore finds that among
the measurements that produced $l=2$, the average value of $k$ can
be much larger than $\avg{k}_{\ket{+}}$ for properly chosen
parameters. This situation leads to the AAV paradox.

\section{Conclusion}

In conclusion, we have presented explanations according to quantum
mechanics of two questions that are relevant to discussions of
weak values. First we presented an example that emphasizes the
role of interpretation in obtaining unphysical results in an AAV
setup. We have also presented the correct interpretation
(according to quantum mechanics) of the measurement results
obtained in an AAV setup. We believe that our discussion is useful
for understanding the origin of the possible observation of
unphysical values in a weak-value experimental setup.

This work was supported in part by the National Security Agency
(NSA), the Laboratory for Physical Sciences (LPS), the Army
Research Office (ARO) and the National Science Foundation (NSF)
grant No.~EIA-0130383.

\end{document}